# Effect of starting materials on the superconducting properties of SmFeAsO$_{1-x}$F$_x$ tapes

Chunlei Wang[1], Chao Yao[1], Xianping Zhang[1], Zhaoshun Gao[1], Dongliang Wang[1], Chengduo Wang[1], He Lin[1], Yanwei Ma[1*], Satoshi Awaji[2], and Kazuo Watanabe[2]

[1] Key Laboratory of Applied Superconductivity, Institute of Electrical Engineering, Chinese Academy of Sciences, P.O. Box 2703, Beijing 100190, China

[2] High Field Laboratory for Superconducting Materials, Institute for Materials Research, Tohoku University, Sendai 980-8577, Japan

**Abstract**

SmFeAsO$_{1-x}$F$_x$ tapes were prepared using three kinds of starting materials. It shows that the starting materials have an obvious effect on the impurity phases in final superconducting tapes. Compared with the other samples, the samples fabricated by SmAs, FeO, Fe$_2$As, and SmF$_3$ have the smallest arsenide impurity phase and voids. As a result, these samples possess much denser structure and better grain connectivity. Moreover, among the three kinds of samples fabricated in this work, this kind of sample has the highest zero-resistivity temperature ~40 K and largest critical current density ~4600 A/cm$^2$ in self-field at 4.2 K. This is the highest $J_c$ values reported so far for SmFeAsO$_{1-x}$F$_x$ wires and tapes.

---

[*] Author to whom any correspondence should be addressed.
E-mail: ywma@mail.iee.ac.cn

# Introduction

The discovery of superconductivity in $SmFeAsO_{1-x}F_x$ with an onset transition temperature $T_c \sim 56$ K has attracted great interest due to its highest $T_c$ in iron-based superconductors and potential applications in high field [1-8]. Compared with the '122' superconductors, the '1111' superconductors have more complex crystal structures and higher sintering temperature [8-10]. Moreover, the F element in the '1111' superconductors is very sensitive to the sintering temperature. A high sintering temperature usually causes serious $F^-$ losses, which will depress the superconducting properties of the $SmFeAsO_{1-x}F_x$ [7].

At a low sintering temperature ($\sim 900^oC$), $SmFeAsO_{1-x}F_x$ bulks with a good superconductivity were prepared by our group [7]. Using one step PIT methods, $SmFeAsO_{1-x}F_x$ superconducting wires and tapes were also successfully synthesized at about 900 $^oC$, achieving a transport $J_c$ of 1.3 KA/cm$^2$ and 2.7 KA/cm$^2$ at 4.2 K, respectively [11-12]. More recently, Fujioka *et al.* reported an ex-situ technology to produce $SmFeAsO_{1-x}F_x$ wire. In order to compensate for the F losses, a binder with stoichiometric Sm, Fe, As, and F were added during the secondary sintering processing, and a $J_c$ about 4 KA/cm$^2$ at 4.2 K was obtained in their works [13]. Generally, the samples fabricated by the above methods have a lot of impurity phases, such as FeAs, SmAs, and SmOF. These impurity phases, especially for arsenide, usually form current-blocking network and reduce the critical current density.

It is well known that the Fe is more prone to react with As to form a stable covalent compound, while Sm tends to form a stable ionic compound with O element. As a result, the reaction between the two stable compounds (FeAs and $Sm_2O_3$) becomes very difficult [14]. Therefore, there are always $Sm_2O_3$ and FeAs impurities in the $SmFeAsO_{1-x}F_x$ compound. However, the impurity phase can be reduced by change the starting materials. In this work, the influence of different starting materials on the superconducting properties of SmFeAsO$_{1-x}$F$_x$ tapes were systematically compared for the first time.

## Experimental details

Fluorine-doped SmFeAsO$_{1-x}$F$_x$ tapes were prepared by three different methods. Sm filings, Fe powder, Fe$_2$O$_3$ powder, As pieces and SmF$_3$ powder were taken as raw materials. Firstly, we synthesized the starting compounds of SmAs, Fe$_2$As, FeO and Sm$_{3-x}$Fe$_{1+2x}$As$_3$ at about 700 $^o$C for 20 hours from the stoichiometric reaction of (Sm + As), (2Fe + As), (Fe$_2$O$_3$ +Fe) and ((1+2x)Fe + (3-x)Sm + 3As, x is determined by the F doping level), respectively. Then, a certain ratio of Sm$_{3-x}$Fe$_{1+2x}$As$_3$, Fe$_2$O$_3$ and SmF$_3$ were mixed together (Named as Sm1111-1), and a certain mass of SmAs, FeO, Fe$_2$As, and SmF$_3$ were blended as the second batch (Named as Sm1111-2). In order to compare with the two kinds of samples mentioned above, a simple one-step PIT method was also adopted (Named as Sm1111-3), following the preparation process in reference [15]. These three mixtures were all thoroughly ground. The final powder was packed into silver tubes with outer- and inner-diameter about 8 mm and 6.2 mm, respectively. Then the sliver tubes were put into iron tubes with outer- and inner-diameter about 11.6 mm and 8.2 mm, respectively. Subsequently, the composite tubes were swaged and drawn down to a wire of ~ 1.9-2.0 mm in diameter. And finally, the wires were rolled as tapes with a thickness ~ 0.6-0.8 mm. Short samples were cut from the as-rolled tapes for sintering. The short tapes were sintered at 500 $^o$C for 10 hour, and then heated at 900 $^o$C for 30-40 hours.

Phase identification was characterized by powder x-ray diffraction (XRD) analysis with Cu-K$\alpha$ radiation from 20$^o$ to 70$^o$. Resistivity measurements were carried out by the standard four-probe method using a PPMS system. The microstructures were determined by scanning electron microscopy (SEM) after peeling away the sheaths. The transport critical currents $I_c$ at 4.2 K and its magnetic dependence were evaluated at the High Field Laboratory for Superconducting Materials (HFLSM) in Sendai, Japan, by a standard four-probe resistive method, with a criterion of 1 $\mu$V cm$^{-1}$. To check the reproducibility, we measured 2–3 specimens for every batch.

**Results and Discussion**

The X-ray diffraction patterns of the three kinds of SmFeAsO$_{1-x}$F$_x$ tapes after heat-treatment are shown in figure 1. Clearly, SmFeAsO$_{1-x}$F$_x$ with ZrCuSiAs structure is the main phase for all kinds of samples. Because it is very difficult to completely clear away the Ag from the surface of SmFeAsO$_{1-x}$F$_x$ core, a tiny amount of Ag reflecting peaks are also observed in the XRD patterns in figure 1. As the fluorine evaporation during the heat-treatment process is unavoidable, especially for these very thin tapes, the final F-doping level will be less than the nominal F-doping level x = 0.2 in present study. According to the figure 1, the main impurity phases for these tapes are SmAs, Sm$_2$O$_3$ and SmOF, and some FeAs phase is also found. However, the relative intensity of the impurity phase is obviously different among these samples using difference starting materials. For the Sm1111-3 samples, all the reflection peaks of the three impurities are very strong and a slight of FeAs impurity is also found, indicating much more impurity in this kind of samples. This is consistent to other results fabricated by the same method [5-6, 11]. For the Sm1111-1 samples, the impurities of Sm$_2$O$_3$ and SmOF were decreased and the main impurity phase is SmAs and FeAs. However, the reflection peaks of SmAs and FeAs were nearly disappeared for the Sm1111-2 samples. The decrease of arsenide are favorable for the good grain linkages, because the arsenide usually exist at the grain boundary of the Sm1111, blocking the superconducting current [16].

The main difference in starting materials for Sm1111-1 and Sm1111-2 samples is $Fe_2O_3$ and FeO. It is obvious that FeO is more active than $Fe_2O_3$. Thus, it is much easier for SmAs to react with FeO (To generate $Sm_2O_3$) than with $Fe_2O_3$. As a result, there are much more $Sm_2O_3$ impurity phases left in Sm1111-2. However the reaction between SmAs and $Fe_2O_3$ become difficult, therefore some SmAs impurity phases are found in Sm1111-1 samples. For Sm1111-3 sample, Sm is a higher electropositive element than that of Fe. Thus Sm can reduce $Fe_2O_3$ to produce $Sm_2O_3$. At the same time, Sm can also react with As to produce SmAs. As a result, the reaction among SmAs, $Sm_2O_3$ and other materials in Sm1111-1 sample becomes very difficult. So there are both $Sm_2O_3$ and SmAs impurity phase left in Sm1111-3 sample. In a word, compared with Sm1111-3 samples, the Sm1111-2 and Sm1111-1 samples are prepared from the metastable compounds, which can effectively decrease the reaction difficulty. Thus the impurities are obviously reduced.

Figure 2 displays the normalized resistivity versus temperature of the three specimens of $SmFeAsO_{1-x}F_x$ tapes. In agreement with the other reports, all specimens show a metallic characteristic before the superconducting transition. The onset superconducting transition temperature and zero resistivity temperature for the Sm1111-1, Sm1111-2 and Sm1111-3 samples are about 46 K, 32 K and 47 K, 40 K and 46 K, 37 K, respectively. It is well known that the $T_c$ for $SmFeAsO_{1-x}F_x$ sample is sensitive to final F-doping level in the crystal, and the obvious $T_c$ suppression is an indicating of the $F^-$ losing. In fact, the loss of $F^-$ is one of the main challenges in preparing $SmFeAsO_{1-x}F_x$ wires and tapes. For example, Wang *et al* prepared

SmFeAsO$_{1-x}$F$_x$ wire with zero resistivity $T_c$ about only 31 K [11], and Masaya *et al* recently reported an Ex-situ technology to fabricate SmFeAsO$_{1-x}$F$_x$ wire with the zero resistivity $T_c$ just about 36 K [13]. The residual resistivity ratio (*RRR*), $R(300)/R(T_c)$ for the Sm1111-1, Sm1111-2 and Sm1111-3 samples are 3.70, 4.47 and 3.57, respectively. The high *RRR* in Sm1111-2 samples indicates that impurity scattering level is low, consistent with the XRD results.

Figure 3 shows the SEM images of SmFeAsO$_{1-x}$F$_x$ tapes, displaying the change of microstructure with different starting materials. It can be seen that there is no obvious difference in grain size and the average grain size is 5-10 $\mu m$. The main difference in the three kinds of specimens is that there are large number of pores in the samples of Sm1111-1 and Sm1111-3. The pore is one of the defects which will reduce the critical current density. In addition, the grain boundary for Sm1111-2 samples seems much cleaner and denser than those of the others, which is consistent with the XRD results. The reduced impurity phases and voids in Sm1111-2 samples can effectively reduce the non-superconducting layer and improve the grain connectivity, resulting in a decrease of current-blocking network. Compared to the other samples, the lesser voids and arsenide impurity phases may be the key factors that make the Sm1111-2 specimens have larger critical current density.

The transport critical current ($I_c$) of the three kinds of SmFeAsO$_{1-x}$F$_x$ tapes were evaluated by the standard four-probe method in fields up to 5 T. The zero resistive current of Sm1111-1, Sm1111-2 and Sm1111-3 samples in self-field is about 12 A, 20 A and 10 A, respectively. The transport critical current density $J_c$ as a function of field for three kinds of tapes is illustrated in figure 4. The largest transport $J_c$ as high as ~ 4600 A/cm$^2$ at 4.2 K in self-field was found in Sm1111-2 samples, and the transport $J_c$ of Sm1111-1 and Sm1111-3 specimens are about 3050 A/cm$^2$ and 2200 A/cm$^2$, respectively. According to the XRD results, the main impurity phase is $Sm_2O_3$ for the Sm1111-2 samples, while there are much more arsenide impurity phases for the Sm1111-1 and Sm1111-3 samples. As reported in literature, the $Sm_2O_3$ phase usually appears in the center of the $SmFeAsO_{1-x}F_x$ grain, whereas the arsenide phases are favor to segregate at the grain boundaries and form grain boundary wetting phase [16]. Therefore, impurity phase of $Sm_2O_3$ is not as harmful as that of arsenide. However, the insulating $Sm_2O_3$ grains together with other defects usually force the current to cross the grain-boundary and restrict current passage regions [16]. As a result, the lowest $J_c$ is found in Sm1111-3 sample. In addition, the Sm1111-2 samples are much denser and should have better grain connectivity than that in Sm1111-1 and Sm1111-3 samples. All these factors are thought to have contributions in the improvement of $J_c$ in Sm1111-2 samples. However, similar to YBCO superconductor, the $J_c$ shows strong field dependence in the low field region for the Sm1111 superconductor, exhibiting an intrinsic weak-link behavior. In the high field region, the transport $J_c$ is nearly field-independent with the highest value ~ 300 $A/cm^2$. So how to overcome the weak-link problem and produce textured or quasi-textured $SmFeAsO_{1-x}F_x$ tapes are the necessary studies for practical applications.

The phase purity improvement of polycrystalline $SmFeAsO_{1-x}F_x$ tapes is a very important work [19]. The Sm1111-2 fabrication process can effectively reduce the arsenide phases and voids, thereby lessening SNS connections between grains and densifying samples. As a result, a $J_c$ value about 4600 $A/cm^2$ was found. This is the highest value reported for $SmFeAsO_{1-x}F_x$ wires and tapes. However, the $J_c$-$B$ performance indicates that the weak-link behavior between grains, which mainly caused by the high-angle misorientation of grains, has no obvious improvement. More recently, our group has achieved progress in *c*-axis aligned $Sr_{0.6}K_{0.4}Fe_2As_2$ superconductor, especially for Sn-doped samples, the transport $J_c$ is as large as 25000 A/cm$^2$ [17-18]. Thus, the methods of texturing and element-doping should be considered in the following works for SmFeAsO$_{1-x}$F$_x$ superconductor.

## Conclusions

The influence of starting materials on the microstructure, impurity phases and transport current density of SmFeAsO$_{1-x}$F$_x$ tapes was investigated. XRD and SEM analysis shows that the impurity phases and voids are sensitive to the starting materials. The highest zero-$T_c$ (~ 40 K) and *RRR* (~ 4.47) were obtained in the sample using SmAs, FeO, Fe$_2$As and SmF$_3$ as starting materials, indicating low impurity scattering level. This is in agreement with the XRD and SEM results. As a result, a high transport $J_c$ about 4600 A/cm$^2$ in self-field was found in these specimens. On the other hand, the transport $J_c$ is very sensitive to applied field and only about 400 A/cm$^2$ left in 0.4 T applied field, showing a weak link behavior.

## Acknowledgements

The authors thank Haihu Wen, Liye Xiao and Liangzhen Lin for their help and useful discussion. This work is partially supported by the National '973' Program (Grant No. 2011CBA00105), National Science Foundation of China (Grant No. 51025726 and 51172230).

# References

[1] Kamihara Y, Watamabe T, Hiramo M and Hosono H 2008 *J. Am. Chem. Soc.* **130** 3296

[2] Chen X H, Wu T, Wu G, Liu R H, Chen H and Fang D F 2008 *Nature* **453** 761

[3] Ren Z A, Lu W, Yang J, Yi W, Shen X L, Zheng C, Che G C, Dong X L, Sun L L, Zhou F and Zhao Zh X 2008 *Chin. Phys. Lett.* **25** 2215

[4] Putti M *et al* 2010 *Supercon. Sci. Technol.* **23** 034003

[5] Gao Z S, Wang L, Qi Y P, Wang D L, Zhang X P, Ma Y W, Yang H and Wen H H 2008 *Supercon. Sci. Technol.* **21** 112001

[6] Zhugadlo N D, Katrych S, Bikowshi Z, Weyeneth S, Puzniak R and Karpinski J 2008 *J. Phys.: Condens. Matter* **20** 342202

[7] Wang C L, Gao Z S, Wang L, Qi Y P, Wang D L, Yao C, Zhang Z Y and Ma Y W 2010 *Supercon. Sci. Technol.* **24** 05002

[8] Ding Y, Sun Y, Wang X D, Zhuang J C, Cui L J and Shi Z X 2011 *Supercon. Sci. Technol.* **24** 095014

[9] Marianne R, Marcus T, and Dirk J 2008 *Phys. Rev. Lett.* **101** 107006.

[10] Guo J G, Jin S F, Wang G, Wang S C, Zhu K X, Zhou T T, He M and Chen X L 2010 *Phys. Rev. B* **82** 180520

[11] Wang L, Qi Y P, Wang D L, Gao Z S, Zhang X P, Wang C L and Ma Y W. 2010 *Supercon. Sci. Technol.* **23** 075005.

[12] Ma Y W, Wang L, Qi Y P, Gao Z S, Wang D L and Zhang X P 2011 *IEEE Trans. Appl. Supercond.* **21** 2878

[13] Fujioka M, Kota T, Matoba M, Ozaki T, Takano Y, Kumakura H, Kamihara Y 2011 *Appl. Phys. Express* **4** 063102.

[14] Fang A H, Huang F Q, Xie X M and Jiang M H 2010 *J. Am. Chem. Soc.* **132**

3260

[15] Ma Y W, Gao Z S, Qi Y P, Zhang X P, Wang L, Zhang Z Y and Wang D L 2009 *Physica C* **469** 651

[16] Kametani F, Li P, Abraimov D, Polyanskii A A, Yamamoto A, Jiang J, Hellstrom E E, Gurevich A, Larbalestier D C, Ren Z A, Yang J, Dong X L, Lu W, and Zhao Z X 2009 *Appl. Phys. Lett.* **95** 142502

[17] Wang L, Qi Y P, Zhang X P, Wang D L, Gao Z S, Wang C L, Yao C and Ma Y W 2011 *Physica C* **471** 1689

[18] Gao Z S, Wang L, Yao C, Qi Y, Wang C, Zhang X P, Wang D L, Wang C D and Ma Y W 2011 *Appl. Phys. Lett.* **99** 242506

[19] Yamamoto A, Jiang J, Kametani F, Polyanskii A, Hellstrom E, Larbalestier D, Martinelli A, Palenzona A, Tropeano M and Putti M 2011 *Supercon. Sci. Technol.* **24** 045010

# Captions

Fig. 1 The XRD patterns for the three tapes produced by different starting materials of (a) $Sm_{3-x}Fe_{1+2x}As_3$, $Fe_2O_3$ and $SmF_3$ (Named as Sm1111-1), (b) SmAs, FeO, $Fe_2As$, and $SmF_3$ (Named as Sm1111-2), and (c) Sm, Fe, As, $Fe_2O_3$ and $SmF_3$ (Named as Sm1111-3).

Fig. 2 The normalized resistivity versus temperature of the three different $SmFeAsO_{1-x}F_x$ tapes labeled as Sm1111-1, Sm1111-2, and Sm1111-3.

Fig. 3 The microstructure of the specimens prepared by different starting materials (a) Sm1111-1, (b) Sm1111-2, (c) Sm1111-3.

Fig. 4 The transport critical current density ($J_c$) versus magnetic field ($\mu_0 H$) for the three different $SmFeAsO_{1-x}F_x$ tapes. The highest $J_c \sim 4.6$ KA/cm$^2$ was obtained in the specimens produced by method of Sm1111-2.

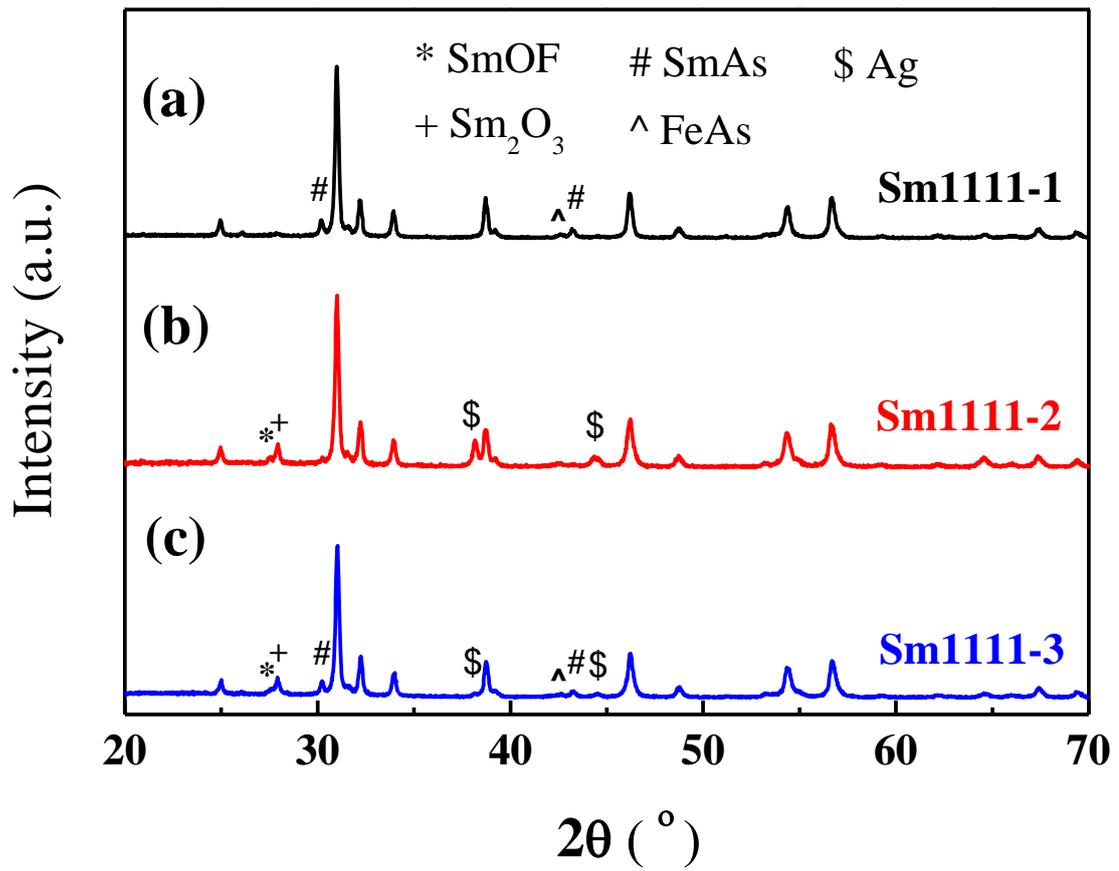

Fig.1 Wang *et al.*

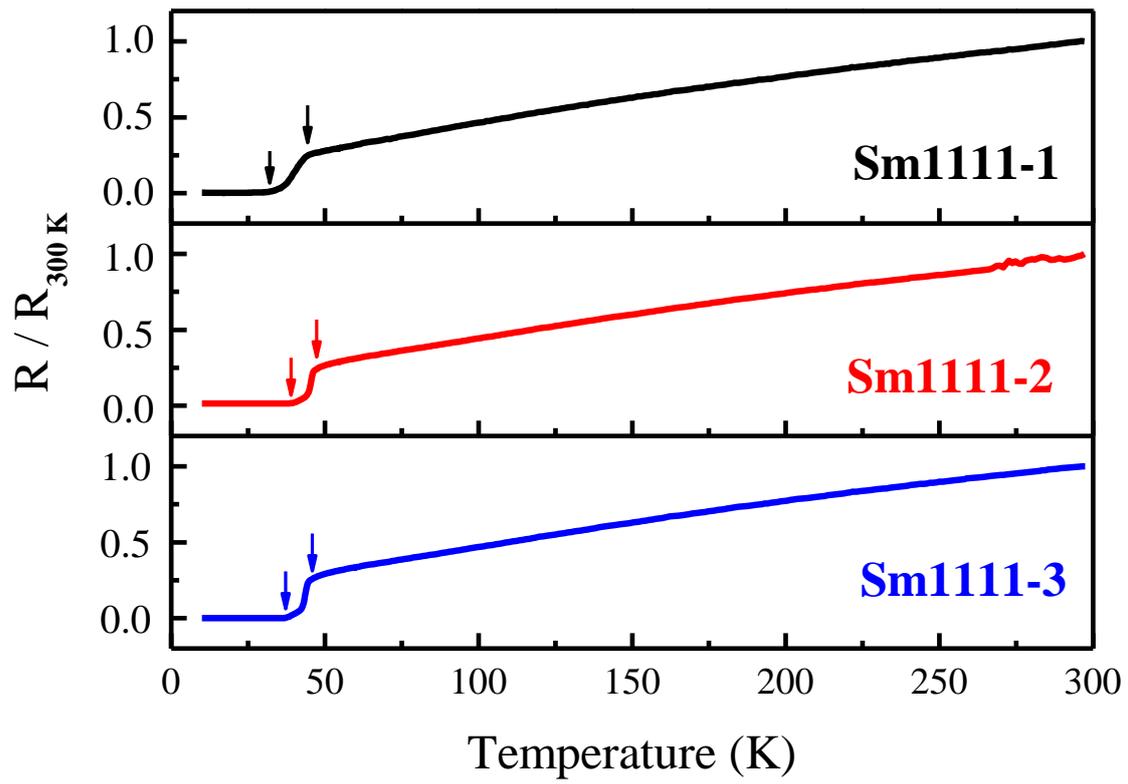

Fig.2 Wang *et al.*

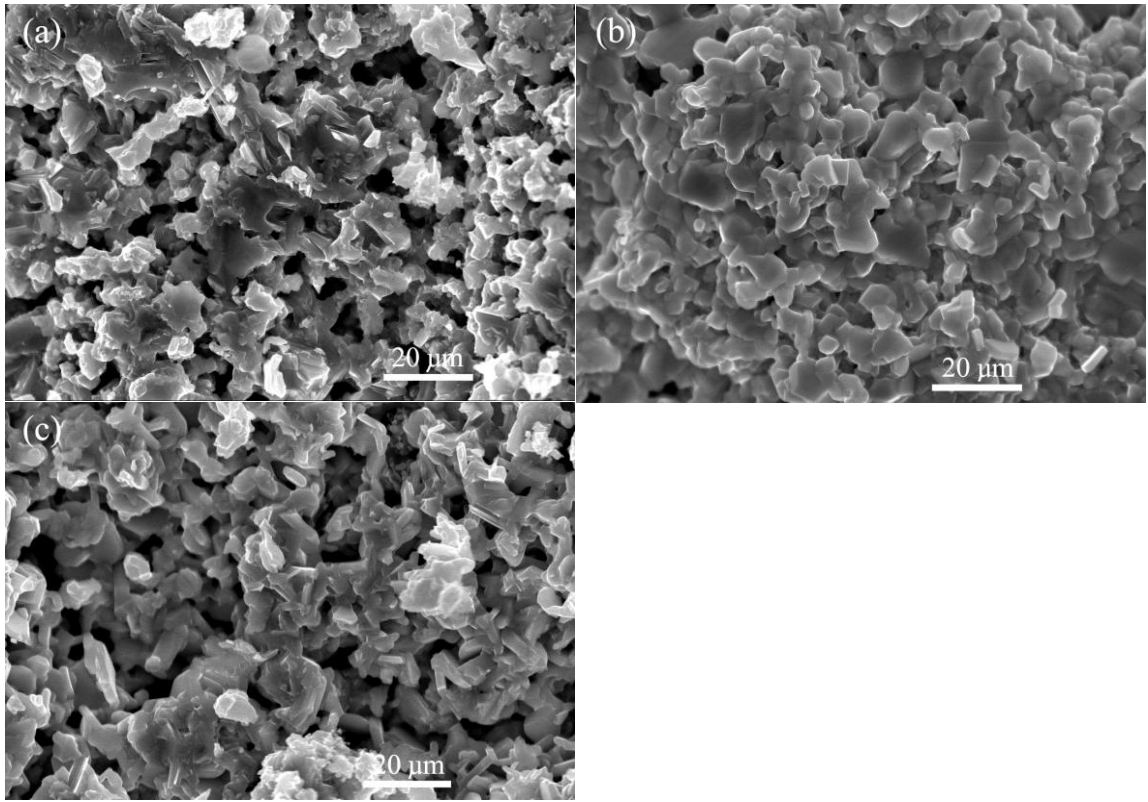

Fig. 3 Wang *et al.*

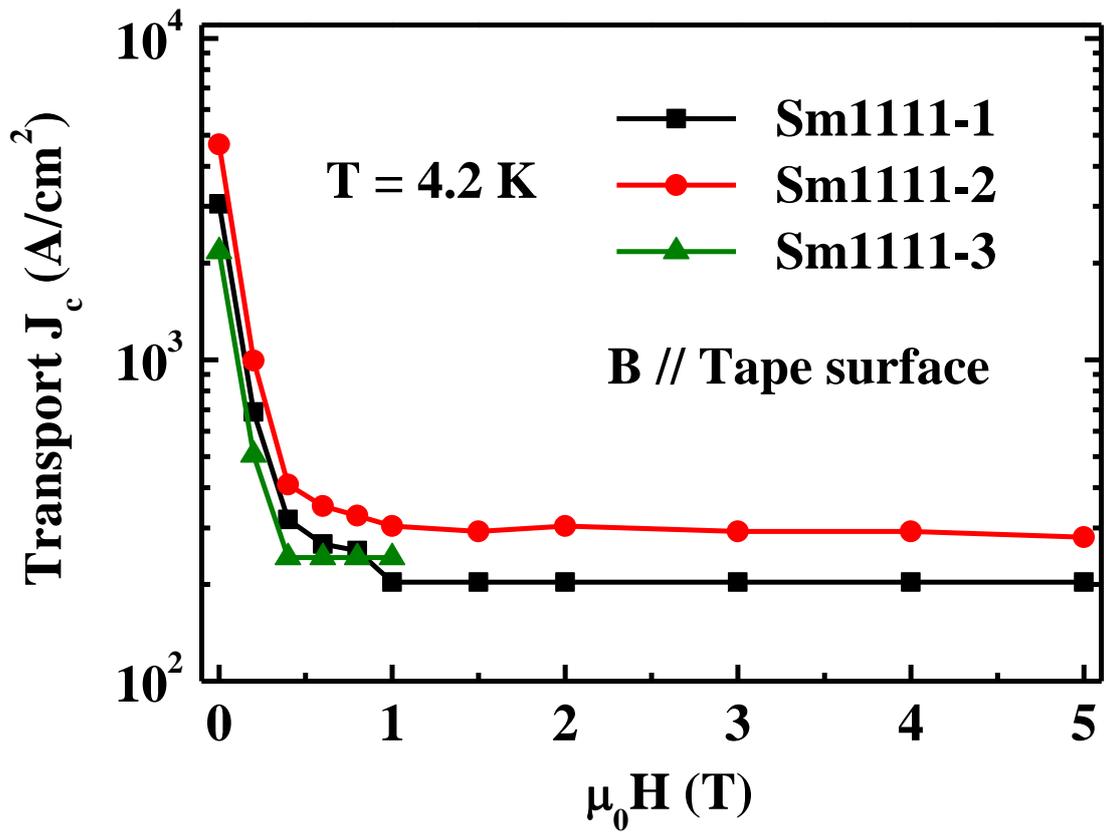

Fig. 4 Wang *et al.*